\documentstyle [emulateapj]{article}
\newcommand{\mgh}[1]{ } 
\newcommand{\mhs}[1]{ }

\newcommand{\op}{Ly$\alpha$\ }

\newcommand{\kms}{\, {\rm km \, s}^{-1} }
\newcommand{\kpc}{\, {\rm kpc} }

\newcommand{\dd}{\,{\rm d}}

\makeatletter

\newenvironment{figurehere}
  {\def\@captype{figure}}
  {}
\makeatother

\begin{document}
\textheight=24.5cm

%
\title{Damped \op absorber and the faint end of the galaxy 
luminosity function\\ at high redshift}

\lefthead{Haehnelt, Steinmetz \& Rauch}
\righthead{Damped \op Absorbers: Observing the Faint End of the Galaxy 
Luminosity Function at High Redshift}

\author{Martin G. Haehnelt\altaffilmark{1},
Matthias Steinmetz\altaffilmark{2,3}, and Michael Rauch\altaffilmark{4}}
\altaffiltext{1}{Max-Planck-Institut f\"ur Astrophysik, Postfach 1523, 85740
Garching, Germany}
\altaffiltext{2}{Steward Observatory, University of Arizona, Tucson} 
\altaffiltext{3}{Alfred P.~Sloan Fellow, David \& Lucile Packard Fellow} 
\altaffiltext{4}{ESO, Karl-Schwarzschild-Str.2, 85740 Garching, Germany }

\centerline{mhaehnelt@mpa-garching.mpg.de, msteinmetz@as.arizona.edu,
mr@eso.org}

\begin{abstract}
We combine predictions for several hierarchical cosmogonies 
with observational evidence on damped \op systems to 
establish a correspondence between the high redshift galaxy 
population and the properties of damped \op  systems (DLAS). We 
assume that high redshift galaxies and damped \op systems 
are hosted by the same dark matter halos  and  require consistency 
between the predicted halo space density, the rate of incidence  
and the velocity width distribution of damped \op systems,
and the observed galaxy luminosity function at the bright end.
 We arrive at the following results: (1) predicted
impact parameters between the damped absorption system and the luminous
parts of the absorbing galaxy are expected to be very small (0.3 - 1
arcsec) for most galaxies; (2) luminosities of galaxies 
causing damped absorption are generally fainter than $m_{_{\rm \cal
R}} = 25$ and 
damped \op systems are predicted to sample preferentially the outer 
regions of galaxies at the faint end of the galaxy luminosity function
at high redshift.  Therefore, DLAS should currently provide  the best
probe of the progenitors of normal present-day galaxies.
\end{abstract}

\keywords{galaxies: kinematics and dynamics ---
galaxies: structure --- quasars: absorption lines} 

\section{Introduction}

The physical conditions inferred from the absorption features 
caused by  high redshift damped \op absorption systems (DLAS) are, 
in several aspects, similar to those in the interstellar medium 
of present-day galaxies. It has therefore been suggested to identify 
DLAS with the progenitors of such galaxies (e.g. Wolfe 1988).
At low redshift  DLAS show a wide variety of morphologies 
(Le Brun et al.\ 1997). At high redshift, however, few have been detected 
in emission (M\o ller \& Warren 1995; Warren \& M\o ller 1995, 
Djorgovski et al.\ 1996,  Djorgovski 1997, M\o ller \& Warren 1998) 
and the nature of  galaxies causing the 
absorption remains unclear for the majority of DLAS at high redshift. 
The main source of information on the nature  of DLAS comes from 
the associated metal absorption which probes the kinematics  
and  the chemical enrichment  history of the mainly neutral gas 
in these systems.   Motivated by the characteristic asymmetric 
shape of the absorption profiles of  low ionization species 
Wolfe and collaborators (e.g. Wolfe 1986) have suggested  
that high-redshift DLAS are large rapidly rotating discs 
akin to present-day spiral galaxies  (but see also Ledoux et al.\ 1998).
In this picture the generally low observed metallicity of high
redshift DLAS (Pettini 1994, Lu 1996, Pettini 1999) is 
taken as evidence that these   are still chemically 
young (Lindner, Fritze-von Alvensleben \& Fricke 1999). 

\begin{deluxetable}{llllllllll}
\tablewidth{0pt}
\tablenum{1}
\tablecaption{\footnotesize Model parameters:
$\sigma_8$ is the {\it rms} linear overdensity in spheres of radius    
$8\,h^{-1}\,{\rm Mpc}$ and $\Gamma$ is a shape parameter 
for CDM-like spectra. h is the Hubble constant
(in  $100\kms$) and $\Omega_0$ and  ${\Omega_{\Lambda}}$   
are the total energy  density and that due to a cosmological constant.
For the other quantities see text.}
\tablehead{
\colhead{${\rm MODEL}$\ \ \ \ }&  
\colhead{${\sigma_8}$\ \ \ \ } &
\colhead{${h}$\ \ \ \  } &
\colhead{${\Omega_0}$\ \ \ \ } &
\colhead{${\Omega_{\Lambda}}$\ \  \ \ }&
\colhead{${\Gamma}$\ \ \ \  }  &
\colhead{$r^{0}_{\rm damp}/\kpc$} &  
\colhead{$r^{0}_{200}/\kpc$} &
\colhead{$m^{0}_{R}$\ \  \ \ }&  
\colhead{$\beta$\ \ \ \  } \\ 
}
\startdata
${\rm SCDM}$ & ${0.67}$ & ${0.5}$ & ${1.0}$ & ${0.0}$ & ${0.5}$&
${16}$&${50}$&${27.1}$&${3}$\\
\\
${\rm \Lambda CDM}$ & ${0.91}$ & ${0.7}$ & ${0.3}$ & ${0.7}$ & ${0.21}$&
${17}$&${64}$&${26.6}$&${2.5}$\\
\\
${\rm OCDM}$ & ${0.85}$ & ${0.7}$ & ${0.3}$ & ${0.0}$ &
${0.21}$&
${16}$&${52}$&${27.4}$&${2.5}$\\ 
\\
${\rm \tau CDM}$ & ${0.67}$ & ${0.5}$ & ${1.0}$ & ${0.0}$ & ${0.21}$&
${26}$&${50}$&${25.9}$&${3}$\\
\enddata
\end{deluxetable} 

The conjecture of DLAS being large rotating discs is, however, at odds
with  the velocity width distribution, the size and the total cross
section of rapidly rotating discs predicted by  hierarchical
cosmogonies (Haehnelt, Steinmetz and Rauch 1998, hereafter HSR98).
Hydrodynamical simulations (Katz et al.\ 1996; Haehnelt, Steinmetz \&
Rauch 1996; Gardner et al.\ 1997ab; Rauch, Haehnelt \& Steinmetz 1997) 
show that gas condensations leading to damped Ly$\alpha$ absorption 
do indeed occur in such scenarios.
However, the hierarchical cosmogonies  predict that a significant
fraction of  the  cross section  for damped absorption is contributed
by gas in halos with circular  velocities  as small as $50 \kms$ (Ma
\& Bertschinger 1994;  Mo \& Miralda-Escud\'e 1994; Klypin et al.\ 1995,
Kauffmann 1996,  Mo, Mao \& White 1999; Gardener et al.\ 1999), 
where we define the circular velocity as $v_{\rm c} =
\sqrt{GM/r}$ at a radius where the overdensity is equal to 200.  In
HSR98 we have shown that the shape of the absorption profiles  is
not unique but can be equally well produced by merging proto-galactic
clumps (cf. Nulsen, Barcon \& Fabian 1998; McDonald \& Miralda-Escud\'e
1999, Maller et al.\ 1999). In HSR98 we further showed that the observed
velocity width distribution of the absorption features of low
ionization species can be reproduced in a standard cold dark matter
(SCDM) cosmology.  At the same time hierarchical cosmogonies such as
SCDM have been shown to reproduce the basic properties of the observed
population of high-redshift galaxies (Adelberger et al.\ 1998;  Steidel
et al.\ 1998; Baugh et al.\ 1998; Bagla 1998; Jing \& Suto 1998; Haehnelt
Natarajan \& Rees 1998; Jing \& Suto 1998;  Contardo, Steinmetz 
\& Fritze-von Alvensleben 1998; Steinmetz 1998; 
Kauffmann et al.\ 1999;  Mo, Mao \& White 1999). Here we
establish a link between  emission and absorption properties of
high-redshift galaxies. In section 2 we briefly test if our previous
findings depend on cosmology. In section 3 we predict the 
luminosity and impact parameter distribution of high-redshift
galaxies responsible for DLAS. Section 4 contains our conclusions.

\section{The velocity width distribution of DLAS}

In hierarchical CDM-like cosmogonies DLAS arise naturally from 
the cool gas that accumulates at the center of dark matter (DM) 
halos (e.g. Katz et al.\ 1996, Gardner et al.\ 1997a/b). 
Absorption features of low ionization species like SiII are generally believed 
to be good tracers of the motion of this gas in a gravitational potential well
(Prochaska \& Wolfe 1997, 1998).  One major characteristic of these 
absorption features is their overall width which
in HSR98 we have shown to be correlated
with the  circular velocity of the DM halo hosting the DLAS. 
However, for a halo of given circular velocity a statistical distribution of 
velocity widths $p((v_{\rm wid}|v_{\rm c})$ arises 
due to  different orientations of the line-of-sight 
and  different dynamical  states  of the gas of the  DM halos.  
Halos with small circular velocities are 
believed to lose most of their gas due to feedback effects. 
The velocity width distribution of  DLAS as probed by low ionization 
species can thus be written as
\begin{equation}
p_{\rm damp}(v_{\rm wid}) = 
\int_{v_{\rm min}} ^{\infty}  
p(v_{\rm wid}| v_{\rm c})\times
p_{\rm damp}(v_{\rm c}) \dd v_{\rm c}  .
\end{equation}
In HSR98 we used simulated absorption profiles for 
a set of haloes with $50\kms < v_{c}< 250 \kms$ to 
investigate $p(v_{\rm wid}|v_{\rm c})$ for a SCDM model and 
found that the distribution does not depend explicitly on 
$v_{\rm c}$ and depends little on redshift. 
We therefore assume $p(v_{\rm wid}| v_{\rm c})$ 
to be a function of $v_{\rm wid}/v_{\rm c}$ only. 
In order to check if there is a dependence on cosmology 
we repeated our analysis for a $\Lambda$CDM model 
(see table 1 for the assumed model parameters and HSR98 for 
details of the simulations).
Figure 1 shows  $p(v_{\rm wid}/v_{\rm c})$ averaged over 
different halos and lines-of-sight. There is little 
difference  between the two cosmological models. 

Once the cross-section weighting for damped 
absorption is known the velocity width distribution is readily 
calculated from  the distribution of 
circular velocities  $p_{\rm damp}(v_{\rm c}) \propto  \sigma_{\rm damp}
\times p_{_{\rm PS}}(v_{\rm c})$. It is currently difficult 
to infer this cross-section weighting reliably from numerical
simulations.  Gardner et al.\ (1997) e.g. found that the cross section 
of DM halos scales as a power-law of  the circular velocity 
as $v_{\rm c}^{2.3-3}$. In their later work they found a much 
shallower scaling $v_{\rm c}^{0.4-1.1}$ (Gardner et al.\ 1999).
This cross-section weighting is likely to be sensitive 
to the energy and momentum input due to supernovae
which is difficult to include properly in  these simulations. 

\begin{figurehere}
\plotone{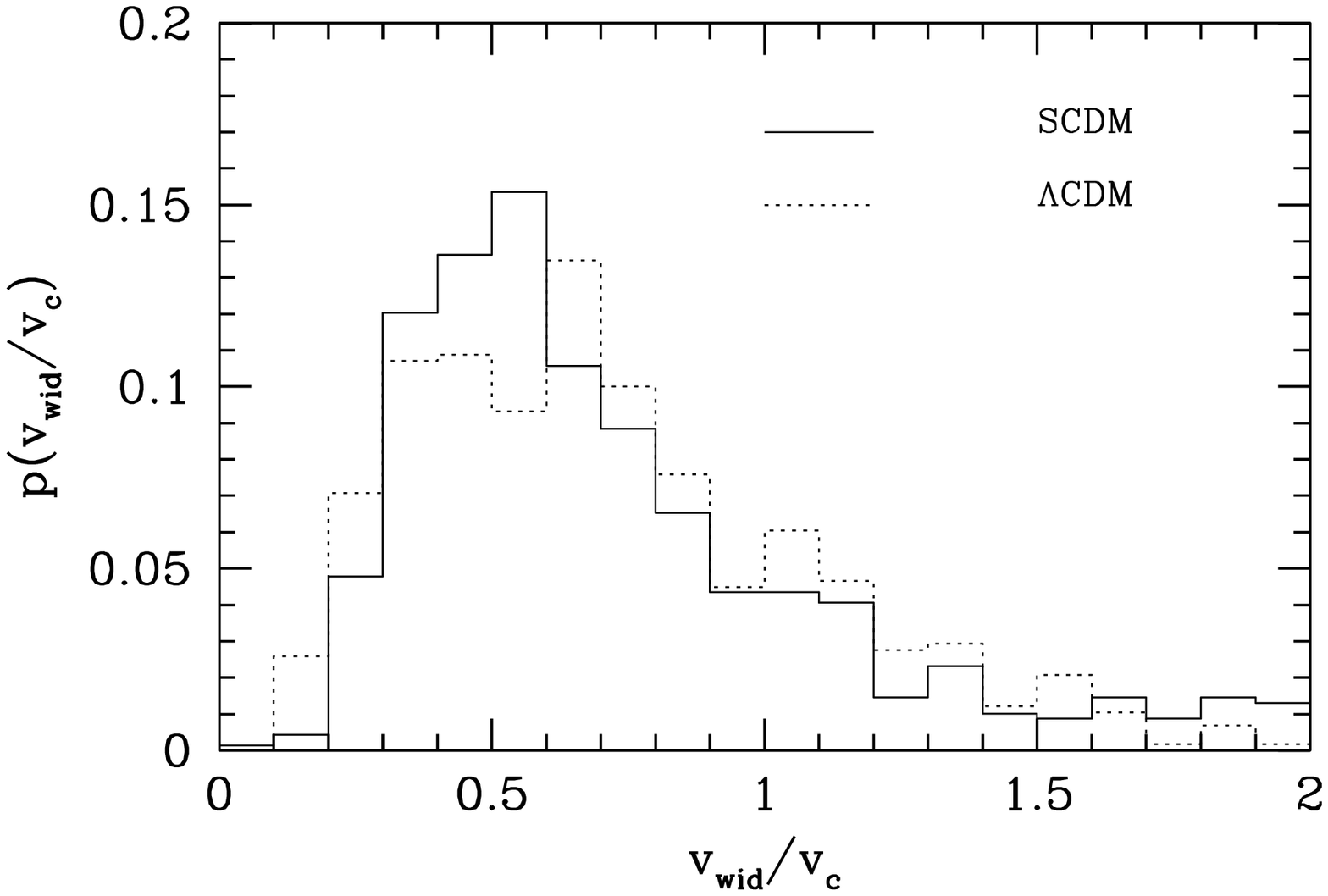}
\caption{\footnotesize  The distribution of velocity 
widths of absorption features of
low ionization species  as a function of $v_{\rm wid}/v_{\rm c}$ 
for two different cosmogonies. Model parameters are given in table 1.} 
\end{figurehere}

\begin{figure*}
\epsscale{2.0} 
\plotone{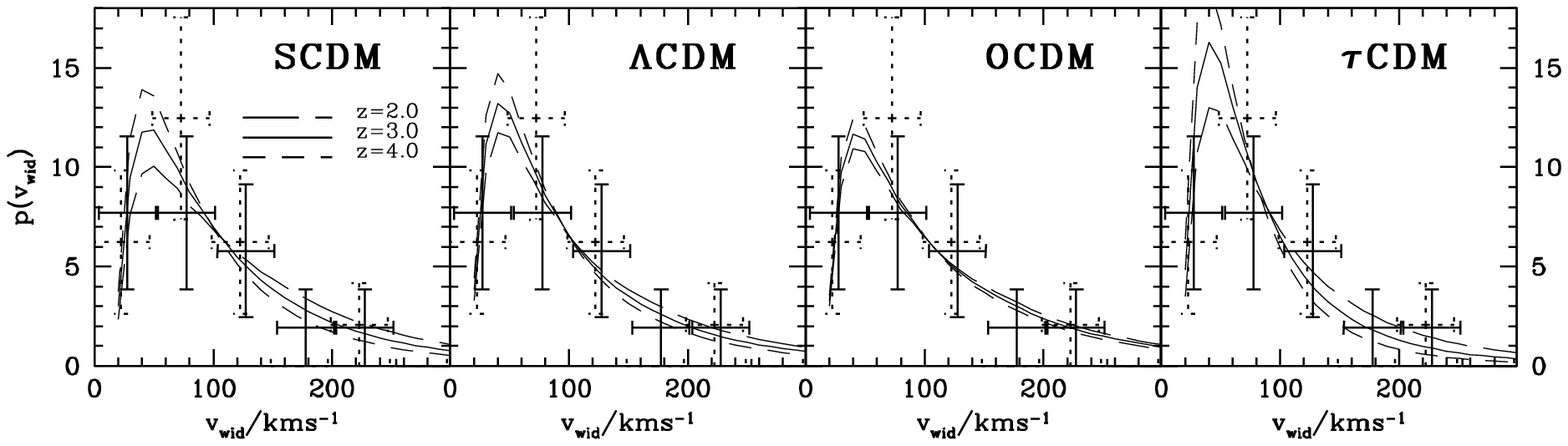}
\caption{\footnotesize The velocity width distribution of the associated 
absorption of low ionization species in DLAS.  The curves 
are model predictions for four different cosmogonies
as calculated from equation (1) (parameters see table 1). 
The crosses are data from Prochaska \& Wolfe (dotted: high z, solid: low z).}
\end{figure*}

\vspace{0.3cm}

Here we take  a different approach and use the observed velocity width 
distribution to determine the mean cross section weighting  for 
damped absorption. We also choose a power-law parameterization 
\begin{equation}
\sigma_{\rm damp} (v_{\rm c}) = \pi (r^{0}_{\rm damp})^2 (v_{\rm c}/200\kms)^\beta. 
\end{equation}
Using the Press-Schechter formalism we calculate 
$p_{_{\rm PS}}(v_{\rm c})$ and obtain the velocity width 
distributions shown in Figure 2. The observed distributions are well 
fit with $\beta \sim 2.5-3$  (see table 1) and $v_{\rm min} = 50
\kms$. Note that there is little evolution with redshift in the observed as 
well as in the predicted distribution. The values of beta are close to
the old value but quite different to the new value of Gardner et al.\ 
(1997,1999). 

\section{Predicting emission propertings}

\subsection{The luminosity distribution of DLAS} 

Recently, it has become possible to determine the  luminosity function for 
the population of star-forming galaxies detected at high redshift
(Steidel et al.\ 1996, Steidel et al.\ 1999). 
The luminosity function at z=3  can be reproduced if a simple 
linear scaling  of the luminosity in the {$\cal R$} band (Steidel et al.\ 1996)
with mass of the DM halo is assumed, 
\begin{equation}
m_{_{\rm \cal R}} =  m_{_{\rm \cal R}}^0 - 7.5\times \log[v_{\rm c}/200\kms]
\end{equation}
(Haehnelt, Natarayan \& Rees 1998). The required  values of 
$m_{_{\rm \cal R}}^{0}$ are  also shown in table 1. 
The same simple model  can also reproduce  the clustering 
strength of high-redshift 
galaxies and its decrease  with decreasing  limiting UV luminosity of 
the galaxy sample (but see Somerville, Primack \& Faber (1999)  
for a somewhat different model). Note that the linear
scaling of the UV luminosity with mass  does not necessarily  imply
that the star formation rate scales linearly with the mass 
of the DM halo as the  dust extinction of the UV luminosity is probably 
luminosity dependent (Steidel et al.\  1999). If we make the plausible
assumption that high-redshift galaxies are hosted by the same population of 
dark matter halos that are causing DLAS  and assume that the linear
scaling of the UV luminosity with mass can be extrapolated 
to smaller masses we can  predict the 
luminosity distribution of DLAS. All we need to know is the scaling of
the cross-section  of DM halos  for damped absorption with circular velocity 
which we already inferred in the last section from the velocity width 
distribution. The result is shown in Figure 3.   80 to 90 percent 
of DLAS are predicted to be fainter than $m_{_{\rm \cal
R}} = 25.5$ (the spectroscopic limit), rather independent of cosmology.  
Some of these may be  faint \op emitters (Fynbo, Thomsen \& M\o ller
1999) but most of them will probably be very difficult to identify 
reliably.

\subsection{The impact parameter distribution of DLAS} 

Similarly we can also 
predict the impact parameter distribution (Fig. 3b). The 
overall rate of incidence, $\dd {\cal{N}}/ \dd  z = 0.2$ 
(Storrie-Lombardi et al.\ 1996),
 determines the normalization of the cross-section
weighting. The corresponding values of $r^0_{\rm
damp} =   r_{\rm  damp} (v_{\rm c} = 200 \kms)$ are  given in table 1.  
The {\em predicted
impact parameters range between  
0.3 and 1 arcsec},  rather independent of cosmology. 
The parameter $r^0_{200} =   r_{200} (v_{\rm c} = 200 \kms)$, i.e.  the 
radius at  which the overdensity is equal to $200 \times \rho_{\rm crit}$
is also listed in table 1. 

The typical  radii of the region causing damped  
absorption are generally about 25 to 50 percent 
of $r_{200}$ of the corresponding DM halo. This is about a 
factor ten larger than the expected scale length of a centrifugally 
supported disc  if the  angular momentum of the gas 
is due to tidal torquing during the collapse of the DM 
halo (Mo, Mao \& White 1998). 
Note that with the assumed scaling of the cross section 
$r_{\rm damp}$ increases more steeply with 
circular  velocity ($\propto v_{\rm c}^{\beta/2}$) 
than $r_{200}$ ($\propto v_{\rm c}$). 

\begin{figure*}
\epsscale{1.0}
\plotone{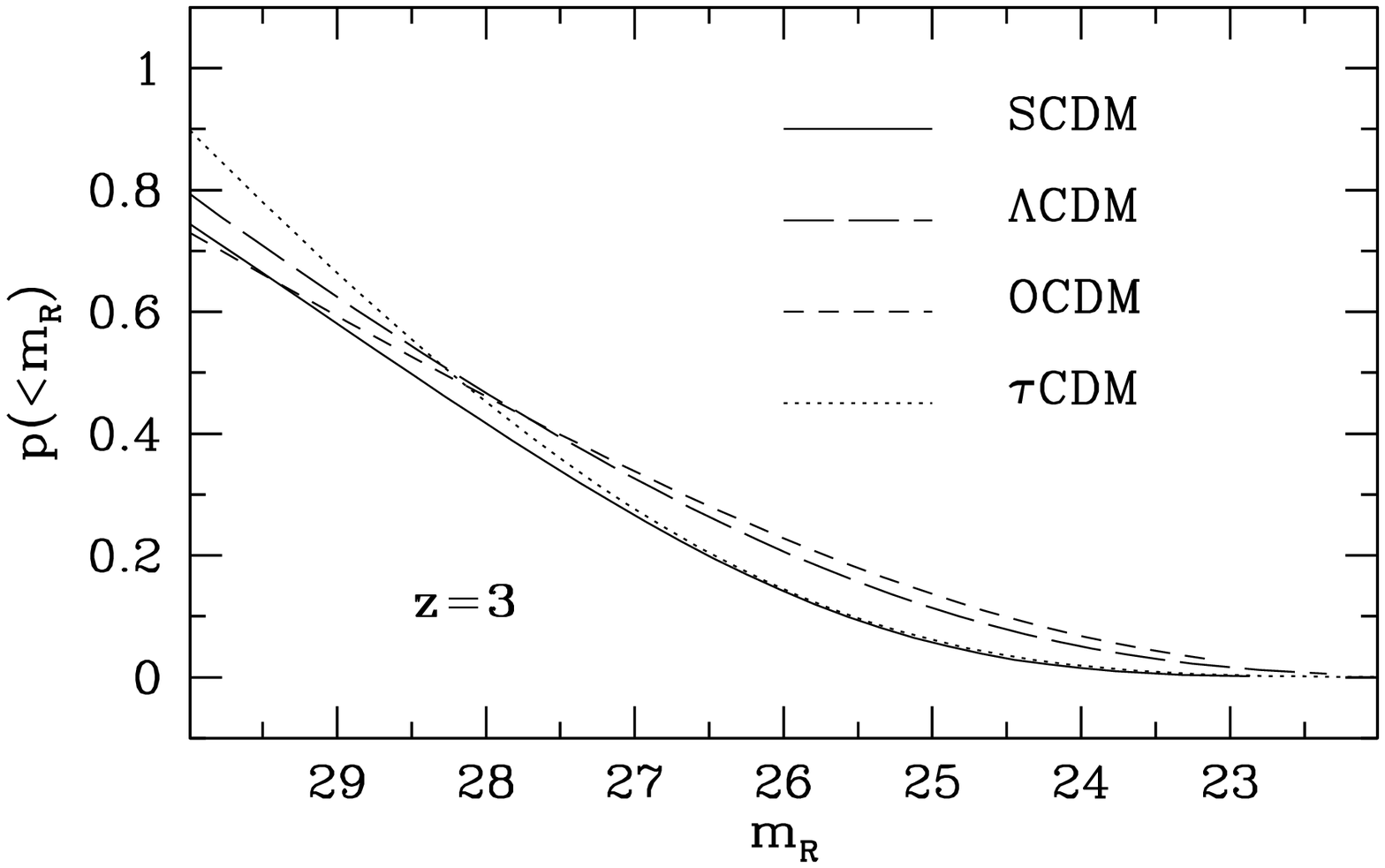}
\plotone{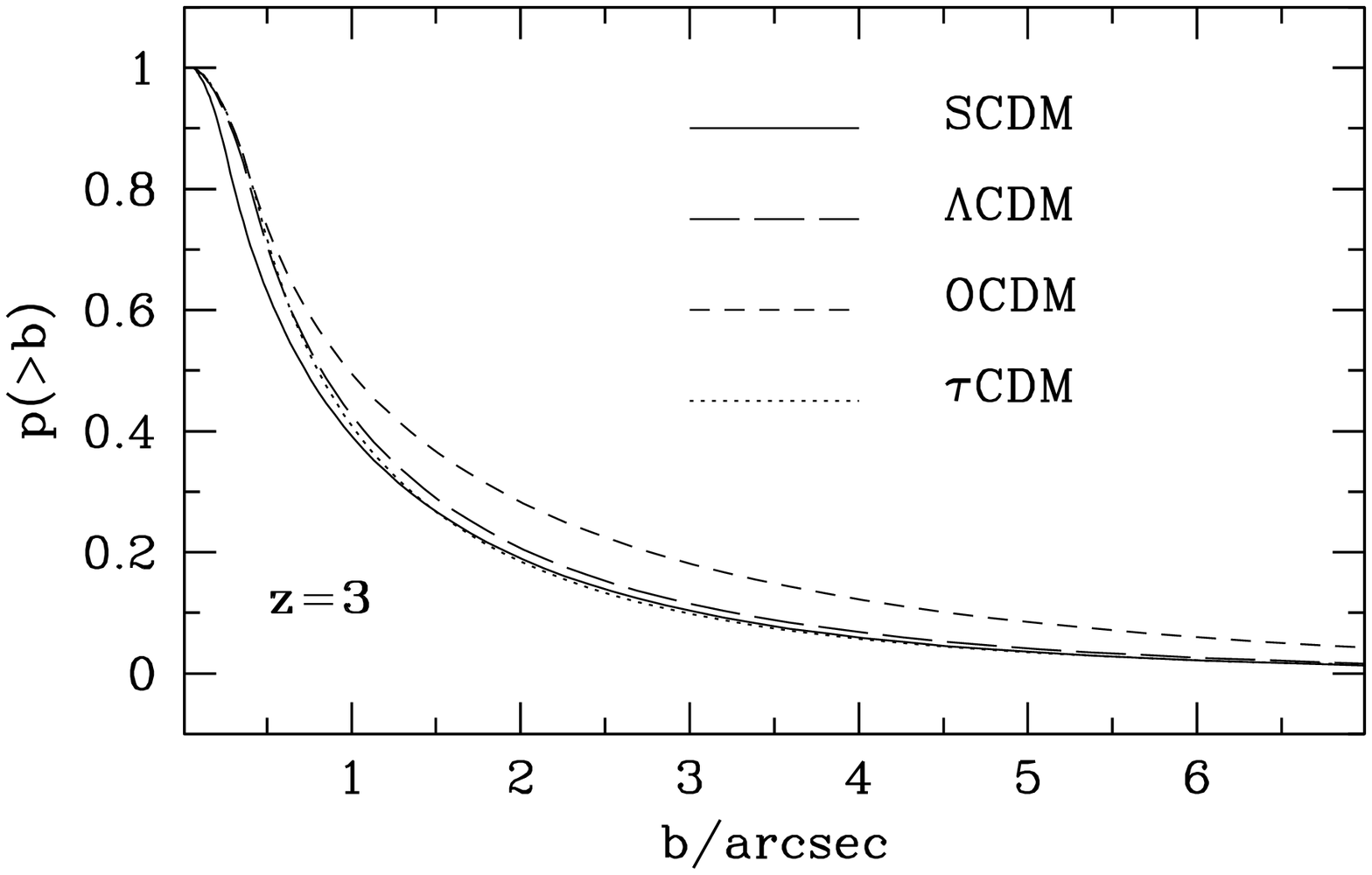}
\caption{\footnotesize Left: The predicted luminosity distribution for 
the DLAS systems for the four different cosmogonies.
Right:The predicted impact parameter distribution.} 
\end{figure*}

\section{Conclusions}

In hierarchical  cosmogonies DLAS are mainly caused by merging 
protogalactic clumps hosted by collapsed DM halos. 
The observed velocity width of the absorption profiles of  
associated low ionization species is well reproduced for a
a range of CDM variants which reproduce the present-day space density of 
galaxy clusters. 

By combining the cross-section weighting inferred from the observed
velocity width distribution with the mass luminosity relation 
inferred  from the luminosity function and clustering properties 
of high redshift galaxies it is possible to link absorption and 
emission properties of DLAS. About 10 to 20\%  of DLAS  are predicted 
to be brighter than $m_{_{\rm \cal R}}=25.5$. Expected impact parameter 
typically range between  0.3 and 1 arcsec but there is a 
pronounced tail of larger impact parameters. 
The predictions seem consistent with the luminosities  
and impact parameters of the small number of DLAS with
spectroscopically identified emission (see e.g. M\o ller 
\&  Warren 1998 for an overview). In our model these would be drawn 
from the tail of the distribution at the bright end and 
at large impact parameter, respectively.
The rather faint flux levels and small  impact parameter 
predicted for the majority of DLAS explains why searches for 
the emission from DLAS have  been notoriously difficult.  

The predicted impact parameters are nevertheless about a factor 
three to five  larger than the typical radii of the  luminous regions of
Lyman-break galaxies. Thus, DLAS at high redshift  should   
preferentially sample the outer regions of galaxies at the faint end  
of the luminosity function. This is consistent with the low 
observed metallicities in DLAS and the outer regions of galaxies. 
In our model DLAS should not be expected to show the high
metallicities estimated for Lyman break galaxies (Pettini
et al.\ 1999);  the latter are derived from emission spectra, which 
are dominated by the light from the more  highly enriched central regions.  

Hierarchical cosmogonies  predict  Lyman-break galaxies 
with $m_{_{\rm \cal R}}<25$ to  become part of bright galaxies  
in galaxy clusters at the present day.  The fainter 
galaxies responsible for DLAS should, however, be the building 
blocks of more typical ($L_{*}$) present-day galaxies consistent 
with  the findings from chemical evolution models
for DLAS. Most high-redshift DLAS are predicted to be too faint to 
obtain  spectra even with  10m telescopes unless they are
strongly magnified by gravitational lensing.  
The analysis of  their absorption properties is currently 
the prime  method for studying the  progenitors of normal
present-day galaxies.

\acknowledgements{We thank Hojun Mo for helpful comments on the manuscript.
This work has been partially supported by 
NATO grant CRG 950752 and by the National Aeronautics and 
Space Administration under NASA grant NAG 5-7151.}


%
\vfill
\end{document}